\documentclass[aps,twocolumn,preprintnumbers,nofootinbib,superscriptaddress]{revtex4-1}
\usepackage{times}
\usepackage{amsmath}
\usepackage{amssymb}
\usepackage{graphicx}
\usepackage{subfigure,accents}
\usepackage{booktabs}
\usepackage{rotating}
\usepackage{longtable}
\usepackage{bm}
\usepackage[pagebackref=false,colorlinks=,linkcolor=blue,citecolor=blue]{hyperref}
\usepackage{epstopdf}
 \pdfoutput=1
 \usepackage[utf8]{inputenc}
\usepackage{cases}
\usepackage{amsmath,float}
\usepackage{amssymb}
\usepackage{amsfonts}
\usepackage{amssymb}
\usepackage{dcolumn}
\usepackage{bm}
\usepackage{bbm}
\usepackage{graphicx}
\usepackage{xcolor}
\usepackage{array}
\usepackage{subfigure}
\usepackage{hyperref}
\usepackage{wasysym}

\newcommand{\be}{\begin{equation}}
\newcommand{\ee}{\end{equation}}
\newcommand{\ba}{\begin{eqnarray}}
\newcommand{\ea}{\end{eqnarray}}

\newcommand{\gsim}{\mathrel{\hbox{\rlap{\lower.55ex \hbox {$\sim$}}
                   \kern-.3em \raise.4ex \hbox{$>$}}}}
\newcommand{\lsim}{\mathrel{\hbox{\rlap{\lower.55ex \hbox {$\sim$}}
                   \kern-.3em \raise.4ex \hbox{$<$}}}}

\hypersetup{colorlinks=true,
            breaklinks=true,
            pdfstartview=Fit,
            linkcolor=blue,
            citecolor=blue,
            urlcolor=blue}
            
 \usepackage{graphicx}

\usepackage{amsmath}

\usepackage{braket}

\newcommand{\bw}{\begin{widetext}}
\newcommand{\ew}{\end{widetext}}

\def\ber{\begin{eqnarray}}
\def\eer{\end{eqnarray}}
\def\beq{\begin{equation}}
\def\eeq{\end{equation}}

\begin{document}

\title{Regular black holes in Verlinde's emergent gravity}
\author{Kimet Jusufi}
\email{kimet.jusufi@unite.edu.mk}
\affiliation{Physics Department, State University of Tetovo, Ilinden Street nn, 
1200,
Tetovo, North Macedonia}
\begin{abstract}
In this paper, we construct a class of charged and regular black hole solutions in Verlinde's Emergent Gravity (VEG). In particular, these solutions include, black holes with asymptotically Minkowski core, T-duality, Frolov/Simpson-Visser type solutions, as well as the Bardeen/Hayward type solutions as a special case. Using the relation between the apparent dark matter and the baryonic matter we find the effect of apparent dark matter on the spacetime geometry. We show that in general, the apparent dark matter leads to non-asymptotically flat spacetime geometry and, in the special limit of a point like mass distribution, all the black hole solutions in VEG resembles the global monopole-like solution with a deficit angle. To this end, we extend the solutions by including the cosmological constant (de Sitter space) and we elaborate different energy conditions of black hole solutions. Finally, having in mind that the Einstein's field equation are not modified in this theory while the apparent dark matter is encoded in the energy-momentum tensor, we used the modified Newman--Janis-Azreg-A\"inou algorithm to obtain the corresponding effective rotating metrics. 
\end{abstract}

\maketitle

\section{Introduction}
Historically, black holes were long considered a mathematical curiosity, however astrophysical observations seems to point out a supermassive black hole in the center of galaxies. In particular, their existence is strongly supported from the recent astrophysical observations reported by the Event Horizon Telescope (EHT) and LIGO. This means that we can test gravity in the strong field regime and get new insights about the structure of spacetime. In the present work, our goal is to investigate the black hole geometry in the context of emergent gravitational theory proposed by Verlinde  \cite{Verlinde:2016toy}. According to this theory, dark matter is an emergent manifestation of gravity which arises from the baryonic mass distribution. More precisely, Verlinde argued that there is an extra gravitational effect due to a volume law contribution to the entropy that is associated with positive dark energy in our universe. In this way, the baryonic mass distribution reduces the entropy content of the universe, this in turn, gives an elastic response of the underlying microscopic system which basically gives the extra gravitational effect, i.e. the dark matter effect as a manifestation of gravity. From the observational point of view, among other things, this theory can explain the rotating flat curve in galaxies. Our main idea in the present work is to use the relation between baryonic and apparent dark matter to construct black hole solutions in VEG and explore the extra effect from the apparent dark matter on the spacetime geometry. 
	 
	The structure of our paper is laid out as follows: In Section II, we construct regular black holes in VEG. In Section III, we study the black hole solutions with non-zero cosmological constant. In Section IV, we study the energy conditions. In Section V, we construct rotating solutions. In Section VI, we comment on our results. 
	
\section{Regular BH solution in VEG}
Accoridng to Verlinde~\cite{Verlinde:2016toy}, by assuming spherical symmetry, then the amount of apparent dark matter $M_D(r)$ is related to the amount of baryonic matter $M_B(r)$  in terms of the following equation
\begin{equation}
 \int_0^r\frac{G M_D(r^\prime)^2}{{r^\prime}^2}dr^\prime = \frac{cH_0M_B(r)r}{6}
\,,
\label{Verlinde's formula}
\end{equation}
where $H_0=2.36\times 10^{-18}~{\rm s^{-1}}\simeq \sqrt{\Lambda/3}$
is the current Hubble parameter, $\Lambda$ is the cosmological
constant
$G=6.674\times 10^{-11}~{\rm m^3/(kg s^2)}$ is the Newton's constant,
$c\approx 3\times 10^8~{\rm m/s}$ is the speed of light. We can also rewrite Eq. (1) as follows 
\begin{equation}
\int_0^r \frac{M^2_D(r') dr}{r'^2}=\frac{a_0 M_B(r) r }{6}
\end{equation}
with $a_0=c H_0$. Furthermore, one can express this relation as
\begin{equation}
M_D^2(r)=\frac{a_0 r^2}{6}\frac{d}{dr}(r M_B(r))
\end{equation}
and define
\begin{equation}
a_M=\frac{a_0}{6},
\end{equation}
where 
\begin{equation}
a_0=5.4 \times 10^{-10} \text{m/s}^2.
\end{equation}

The simplest scenario is to consider a a spherically symmetric black hole solutions with the line element
\begin{equation}
\label{eq:lineElem}
ds^2
= g_{tt}dt^2 +g_{rr}dr^2 +r^2(d\theta^2+\sin^2\theta d\phi^2).
\end{equation}
For the black hole case, we can assume $g_{tt}=-g_{rr}^{-1}$ and 
\begin{eqnarray}
    g^{-1}_{rr}=1-\frac{2 m(r)}{r},
\end{eqnarray}
where the mass profile is given by
\begin{eqnarray}
    m(r)=4 \pi \int_0^r [\rho_B(r')+\rho_D(r')] r'^2 dr'.
\end{eqnarray}

In what follows we are going to construct regular black holes with different mass profiles and see the effect of the apparent dark matter on the spacetime geometry.

\begin{figure*}{}
    \centering
        \includegraphics[scale=.42]{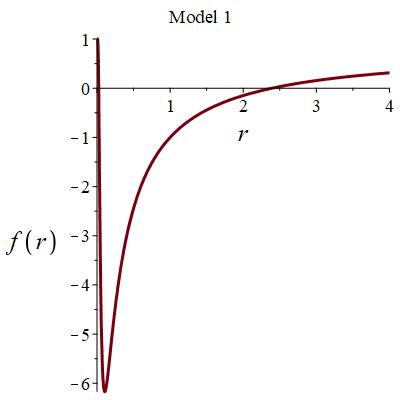}
          \includegraphics[scale=.42]{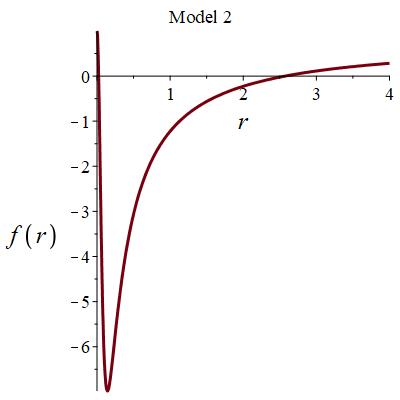}
          \includegraphics[scale=.42]{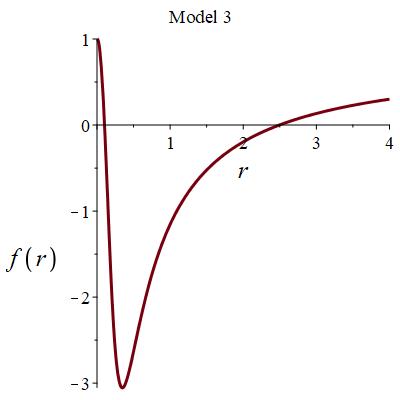}
           \includegraphics[scale=.42]{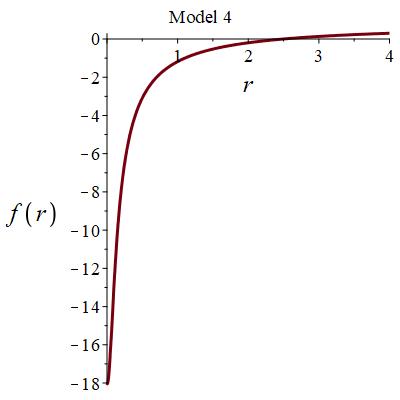}
       \caption{The plot for the metric function $f(r)$ using $l_0=0.1, M=1, a_M=0.01,$ and $Q=0.1$. }\label{f0}
\end{figure*}
       
\subsection{Model 1 (Charged Minkowski core)}
Our first example is to consider the mass profile proposed by Simpson and Visser given by the exponentially decreasing law for the baryonic matter, i.e., $ M_B(r)=M \exp{(-l_0/r^n)}$ \cite{Simpson:2019mud}. Here we shall modify this profile by adding an electric charge as follows
\begin{eqnarray}
    M_B(r)=\left(M-\frac{Q^2}{2r}\right) \exp{(-l_0/r^n)}.
\end{eqnarray}
Solving for the apparent dark matter first and then using $f(r)=-g_{tt}=g_{rr}^{-1}$, we can find the general solution for the metric function
\begin{equation}
    f(r)=1-\left(\frac{2 M}{r}-\frac{Q^2}{r^2}\right)e^{-\frac{l_0}{r^n}}-\frac{2\sqrt{a_M} e^{-\frac{l_0}{2r^n}}}{r} \mathcal{A}(r)
\end{equation}
where we have introduced
\begin{eqnarray}
\mathcal{A}=\sqrt{r \left(Mr+\frac{M l_0 n}{r^{n-1}}-\frac{Q^2 l_0 n }{2 r^n}\right)}.
\end{eqnarray}

This metric is everywhere regular and the scalar curvature invariants such as the Ricci scalar and the Kretschmann scalar are finite when $r\to 0$, i.e.,
\begin{eqnarray}
\lim_{r \to 0}R_{\mu \nu}R^{\mu \nu}\to 0,\,\,\,\text{and}\,\,\,\,
\lim_{r \to 0}R_{\mu \nu \sigma \eta}R^{\mu \nu \sigma \eta}\to 0.
\end{eqnarray}
In Fig. 1, we show the location of the outer and inner horizon for this black hole for specific values of parameters. 

\subsubsection{Special case: Simpson-Visser Minkowski core solution in VEG}
A special case follows if we set the charge to zero, i.e., $Q=0$, this yields
\begin{equation}
    f(r)=1-\frac{2 M}{r} e^{-\frac{l_0}{r^n}}-\xi e^{-\frac{l_0}{2r^n}}\sqrt{1+\frac{ l_0\, n}{r^{n}}}.
\end{equation}
where we defined the quantity $\xi=2 \sqrt{a_M M}$. This solution is still regular. Furthermore, in the limit $l_0=0$, from the last equation we obtain 
\begin{eqnarray}
  f(r)=1-\frac{2M}{r}-\xi.
\end{eqnarray}
 Quite interestingly, this equations resembles precisely the well known global monopole solution  \cite{Barriola:1989hx}. The fact that Verlinde's theory admits a global monopole-like solution was shown in \cite{lei} where the case of a relativisitc star was explored. In other words, due to the gravitational collapse of the baryonic matter the resulting black hole spacetime will have a deficit angle arising from the effect of apparent dark matter. We point out also that in Ref. \cite{Jusufi:2020zln} the spacetime of the Schwarzschild black hole with surrounding baryonic matter was investigated.  In that work, the black hole was asssumed to be Schwarzschild black solution and the effect of the deficit angle was neglected, instead the effect of the surrounding matter on the black hole shadow was analysed.   

\subsection{Model 2 (T-duality)}
The concept of T-duality is known in string theory and basically identifies theories on higher-dimensional spacetimes with mutually inverse compactification radii~$R$. In particular one can use the correspondence 
$R \rightarrow {R^\star}^2/R$ and $n \leftrightarrow w$, where $R^\star = \sqrt{\alpha'}$ is known as the self-dual radius, in addition $\alpha'$ represents the Regge slope, furthermore $n$ gives the Kaluza-Klein excitation and $w$ gives the winding number  \cite{Nicolini1}. From the work of Padmanabhan on the concept of T-duality, it was shown that the gravitational potential and electric potential are modified due to the finite length $l_0$. In particular, for the potential of the system which consists of two pointlike masses, $m$ and $M$, at relative distance $\vec{r}$, one can write \cite{Nicolini1}
\begin{align}
\phi(r) \to -\frac{M}{\sqrt{r^2 + l_0^2}}.
\label{eq:statpot}
\end{align}
For the electric potential on the other hand it was shown \cite{Nicolini3,Nicolini2}
 \begin{eqnarray}
  V(r)=-\frac{Q}{\sqrt{r^2+l_0^2}}.
\end{eqnarray} 

This leads to the following mass function (see for details \cite{Nicolini2})
\begin{eqnarray}
M_B(r)=\frac{M\,r^3}{(r^2+l_0^2)^{3/2}}+4\pi \int_0^r \frac{Q^2 r'^4}{8 \pi (r'^2+l_0^2)^3}dr'
\end{eqnarray}
we obtain 
\begin{equation}
    f(r)= 1-\frac{2M r^2}{\left(r^2+l_0^2\right)^{3/2}}+\frac{Q^2 r^2 \mathcal{H}(r)}{(r^2+l_0^2)^2} -2\sqrt{a_M\mathcal{G}(r)},
\end{equation} 
where 
\begin{eqnarray}
 \mathcal{H}(r)=\frac{5}{8}+\frac{3 l_0^2}{8 r^2}
 -\frac{3 (r^2+l_0^2)^2 }{8 l_0 r^3}\arctan\left(\frac{r}{l_0}\right),
\end{eqnarray}
and 
\begin{eqnarray}\notag
\mathcal{G}(r)&=&\frac{3 Q^2 }{16 l_0 }\arctan\left(\frac{r}{l_0}\right)+\frac{M r^3(4 l_0^2+r^2)}{(l_0^2+r^2)^{5/2}}\\
&-&\frac{Q^2 r(3 l_0^4+8 l_0^2 r^2 -3 r^4)}{16(r^2+l_0^2)^3}.
\end{eqnarray}
This metric is everywhere regular with the scalar curvature invariants such as the Ricci scalar and the Kretschmann scalar are finite when $r\to 0$.  It is also important to stress out that in solution (18-20), $M$ is the mass parameter and not the true mass of the system measured at infinity. For example, by neglecting the effect of apparent dark matter for a moment, one can show that the ADM mass $\mathcal{M}$ is related to the mass parameter $M$ by \cite{Nicolini2}
\begin{eqnarray}
    \mathcal{M}=M+\frac{3 \pi Q^2}{32 \pi l_0}
\end{eqnarray}
which physically means that the the mass contains a new term proportional to the regularized self energy of the electrostatic field. Using the last relation we can also express the solution (18) in terms of the ADM mass, however, in the present work due to the apparent dark matter the solution in general is not asymptotically flat and we must consider a cut off or finite distance corrections. In Fig. 1, we have shown the location of the outer and inner horizon for this black hole.

 \subsubsection{Special case: Bardeen type solution in VEG}
One can now easily see that we can recover the Bardeen solution as a special case by setting $Q=0$. It has the mass profile \cite{Bardeen}
\begin{eqnarray}
    M_B(r)=\frac{M\,r^3}{(r^2+l_0^2)^{3/2}},
\end{eqnarray}
provided $l_0=2M L^2$. Solving for the apparent dark matter we obtain
\begin{eqnarray}
    f(r)=1-\frac{2 M\,r^2}{(r^2+l_0^2)^{3/2}}-\xi \sqrt{\frac{r^3\, (4 l_0^2+r^2)}{(r^2+l_0^2)^{5/2}}}
\end{eqnarray}

Unlike the case of Minkowski core, here  it is very interesting to note that for $Q=0$, one can show that the Kretschmann scalar and the Ricci scalar goes to infinity as $r \to 0$, i.e.,  Ricci scalar and the Kretschmann scalar diverge as $r\to 0$. This is a surprising result since in the limit $\xi=0$ we obtain the Bardeen solution in GR which is indeed a regular solution.  However, even thought there is a singularity for solution (23) compared to the point like mass or the classical Schwarzschild black hole, due to the apparent dark matter effect, light rays (and particles) never reach  the singularity. Let us consider a reference frame of an observer falling from rest towards the black hole known as the Painlev\'e-Gullstrand coordinates. Introducing $dT=dt-h(r) dr$, in the equatorial plane ($\theta=\pi/2$), after fixing $h(r)=\sqrt{1-f(r)}/f(r)$, we obtain the metric
\begin{eqnarray}
    ds^2=-f(r)dT^2+2 \sqrt{1-f(r)}dTdr+dr^2,
\end{eqnarray}
which is regular at the black hole horizon. If we solve for radial light geodesics $ds^2=0$, one can find two solutions
\begin{eqnarray}
    \frac{dr}{dT}=-\sqrt{\frac{2 M\,r^2}{(r^2+l_0^2)^{3/2}}+\xi   \sqrt{\frac{r^3\, (4 l_0^2+r^2)}{(r^2+l_0^2)^{5/2}}}} \pm 1,
\end{eqnarray}
in which we have two situations: for positive sign the light is moving in the outward direction; for the negative sign the light is moving in the inward direction. We know for the Schwarzschild black hole that the quantity $dr/dT$ of both the inward and outward light rays go to negative infinity as $r \to 0$, this physically means that anything that enters the event horizon of the black hole reaches the singularity at the center. However, in the case of Bardeen solution in VEG, we can see from Fig. 2 that $dr/dT$ reaches a minimum value at some $r$ and then increases for smaller values of $r$.  This also means that, below this value, the quantity $dr/dT$ is imaginary, which means that nothing can reach the center, and the
observed curvature scalars should remain finite. Such a singularity is known as a timelike singularity, which is different from the Schwarzschild black hole which harbors a spacelike singularity. A similar situation has been studied in Refs. \cite{Ali:2015tva,Jusufi:2022uhk} for quantum and string corrected black holes.  
\begin{figure*}
    \centering
        \includegraphics[scale=.43]{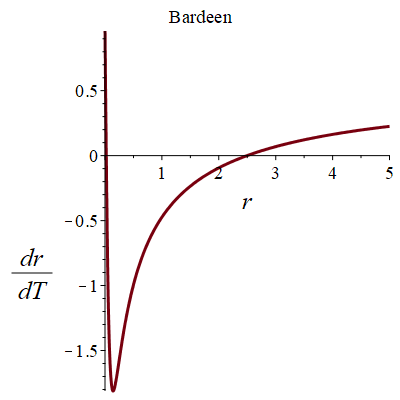}
          \includegraphics[scale=.43]{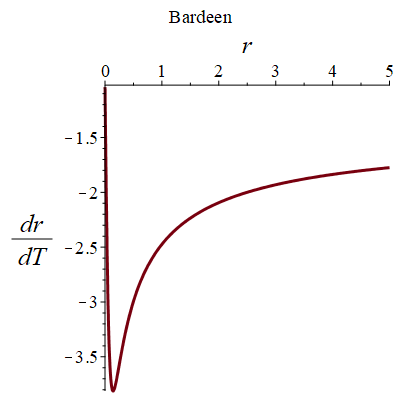}
           \includegraphics[scale=.43]{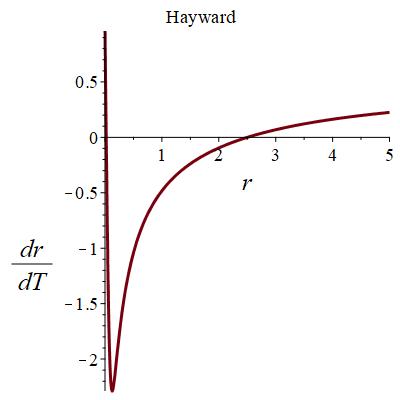}
          \includegraphics[scale=.43]{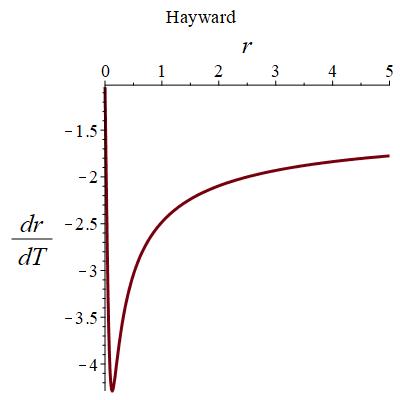}
       \caption{The plot shows the velocity of light observed in a freely falling frame for light moving inward. Right panel: The velocity of light observed in a freely falling frame for light moving outward  using $l_0=0.1, M=1, a_M=0.01.$ }\label{f1}
\end{figure*}

\subsection{Model 3 (Frolov type)}
Next, we follow Frolov where the following mass function as introduced \cite{Frolov}
\begin{eqnarray}
M_B(r)=\frac{(Mr-\frac{Q^2}{2})r^3}{r^4+(2Mr+Q^2)l_0^2}
\end{eqnarray}
which leads to the following solution 
\begin{equation}
    f(r)=1-\frac{(2Mr-Q^2)r^2}{r^4+(2Mr+Q^2)l_0^2}-2 \frac{\sqrt{a_M \mathcal{K}(r)}}{r^4+(2Mr+Q^2)l_0^2},
\end{equation}
with
\begin{eqnarray}
\mathcal{K}=r^3(Mr^5+l_0^2(8M^2r^2+2M Q^2r-2 Q^4)).
\end{eqnarray}

This spacetime is regular as the Ricci scalar and the Kretschmann scalar are finite when $r\to 0$. In Fig. 1, we can see the location of the outer and inner horizon for this black hole for specific values of parameters.

\subsubsection{Special case: Hayward type solution in VEG}
If we set the charge to zero $Q=0$, it can be obtain the Hayward mass function which has the mass profile \cite{Haward}
\begin{eqnarray}
    M_B(r)=\frac{M\,r^3}{r^3+l_0^3},
\end{eqnarray}
having used the rescaling $l_0^3 \to 2M l_0^2$. Solving for the apparent dark matter we obtain the solution
\begin{eqnarray}
    f(r)=1-\frac{2 M\,r^2}{r^3+l_0^3}-\xi \frac{\sqrt{r^3\,(4l_0^3+r^3)}}{r^3+l_0^3}
\end{eqnarray}
We see that the Hayward black hole solution is modified due to the extra term in the metric function as a result of apparent dark matter. In the limit $l_0=0$, again we obtain the global monopole solution. Quite similar to the  case of Bardeen solution, the Hayward solution in VEG has scalar invariants which diverge as $r\to 0$. However, if we consider a freely falling frame with
\begin{eqnarray}
    \frac{dr}{dT}=-\sqrt{\frac{2 M\,r^2}{r^3+l_0^3}+\xi \frac{\sqrt{r^3\,(4l_0^3+r^3)}}{r^3+l_0^3}} \pm 1,
\end{eqnarray}
one can see from Fig. 2 that the velocity reaches a minimum value at some $r$ but then increases for smaller values of $r$, and this means that nothing can reach the center. Such a singularity as we already pointed out is known as a timelike singularity.

\subsection{Model 4 (Simpson-Visser type)}
In our last model, we follow \cite{Simpson:2018tsi} where the metric the spacetime geometry is given by
\begin{equation}
\label{eq:lineElem}
ds^2
= g_{tt}dt^2 +g_{rr}dr^2 +(r^2+l_0^2)(d\theta^2+\sin^2\theta d\phi^2).
\end{equation}
 It has the mass profile \cite{Simpson:2018tsi}
\begin{eqnarray}
    M_B(r)=\frac{Mr}{\sqrt{r^2+l_0^2}}-\frac{Q^2r}{2(r^2+l_0^2)}
\end{eqnarray}
yielding the solution 
\begin{equation}
f(r)=1-\frac{2M}{\sqrt{r^2+l_0^2}}+\frac{Q^2}{r^2+l_0^2}-2 \sqrt{ \frac{a_M r \mathcal{R}(r)}{(r^2+l_0^2)^{5/2}}}
\end{equation}
where 
\begin{eqnarray}
\mathcal{R}=(r^2+l_0^2)(r^2+2l_0^2)M-Q^2 l_0^2 \sqrt{r^2+l_0^2}.
\end{eqnarray}
The solution describes a regular spacetime geometry.  Unlike the case investigated in \cite{Simpson:2018tsi}, here, due to the apparent dark matter correction term proportional $a_M$, for the region $r<0$, we obtain complex metric, hence we shall consider only the region $r \geq 0$, as a physically accepted region.   Depending on the specific values for parameters,  this metric can describe not only a black hole but also a wormhole solution. In Fig. 1, we show the location of the black hole horizons for this model.

\subsubsection{Special case: Simpson-Visser solution in VEG}
Taking the charge to be zero, i.e. $Q=0$, we obtain the conical spacetime
\begin{equation}
f(r)=1-\frac{2M}{\sqrt{r^2+l_0^2}}-\xi \sqrt{\frac{r (r^2+2l_0^2) }{(r^2+l_0^2)^{3/2}}}.
\end{equation}
provided $r\geq 0$ for the same reasons we have explained above. This solution generalises the  black-bounce spacetime geometry studied in \cite{Simpson:2018tsi}. The difference is that, due to the apparent dark matter effect the spacetime is not asymptotically flat. For these reasons,  we refer to $M$ as the mass parameter, not the true ADM mass of the system. If we further set $l_0 \to 0$, then the above solution resembles the point like mass with conical geometry described described by Eq. (14).
Finally, we shall consider one more special case when when $M\to 0$ and $Q\to 0$, in that case we get the Morris-Thorne wormhole spacetime 
\begin{eqnarray}
ds^2=-dt^2+dr^2+(r^2+l_0^2)(d\theta^2+\sin^2\theta d\phi^2).
\end{eqnarray}
We can see that the effect of apparent dark matter vanishes when we set the matter field to zero. This is to be expected due to the relation between them given by Eq. (3). The interesting conclusion is that the effect of apparent dark matter in VEG leads to non-asymptotically flat spacetime geometries. 

\section{Black holes in de Sitter space in VEG}
The solutions found in the previous section can be generalized by including the cosmological constant. The Einstein field equations in VEG can be written as 
\begin{eqnarray}
R_{\mu\nu}-\frac{1}{2}g_{\mu \nu}R=8\pi T_{\mu\nu},
\end{eqnarray}
where $T^{\mu}_{\nu}=(-\tilde{\rho},P_r,T_t,P_t)$ and $\tilde{\rho}=\rho_B+\rho_D+\rho_\Lambda$, since we have moved the cosmological constant in the right hand side and the mass profile is now given by
\begin{eqnarray}
    m(r)=4 \pi \int_0^r [\rho_B(r')+\rho_D(r')+\rho_\Lambda(r')] r'^2 dr'.
\end{eqnarray}
This means we have to take into account the baryonic part,the apparent dark matter contribution, and the cosmological constant contribution. Let us point out that, even if we neglect the presence of cosmological constant for e moment in the Einstein field equations, the presence of dark energy is still fundamental in VEG - its effect is encoded in the apparent dark matter term which is proportional to $a_M\sim \sqrt{\Lambda/3}.$  By including the cosmological constant as part of the energy-momentum tensor, we have an extra term $\Lambda r^2/3$, with the corresponding solutions:
	\begin{itemize}
	    \item Model 1
	\end{itemize}
	\begin{equation}
    f(r)=1-\left(\frac{2 M}{r}-\frac{Q^2}{r^2}\right)e^{-\frac{l_0}{r^n}}-\frac{2\sqrt{a_M} e^{-\frac{l_0}{2r^n}}\mathcal{A}}{r} -\frac{\Lambda r^2}{3},
\end{equation}
\begin{itemize}
	    \item Model 2
	\end{itemize}
	\begin{equation}
    f(r)= 1-\frac{2M r^2}{\left(r^2+l_0^2\right)^{3/2}}+\frac{Q^2 r^2 \mathcal{H}}{(r^2+l_0^2)^2} -2\sqrt{a_M\mathcal{G}} -\frac{\Lambda r^2}{3},
\end{equation}
	\begin{itemize}
	    \item Model 3
	\end{itemize}
\begin{equation}
    f(r)=1-\frac{(2Mr-Q^2)r^2}{r^4+(2Mr+Q^2)l_0^2}- \frac{2\sqrt{a_M \mathcal{K}}}{r^4+(2Mr+Q^2)l_0^2}-\frac{\Lambda r^2}{3},
\end{equation}
	\begin{itemize}
	    \item Model 4
	\end{itemize}
\begin{equation}
f(r)=1-\frac{2M}{\sqrt{r^2+l_0^2}}+\frac{Q^2}{r^2+l_0^2}-2 \sqrt{ \frac{a_M r \mathcal{R}}{(r^2+l_0^2)^{5/2}}}
-\frac{\Lambda r^2}{3}. 
\end{equation}

\begin{figure*}{}
        \includegraphics[scale=.42]{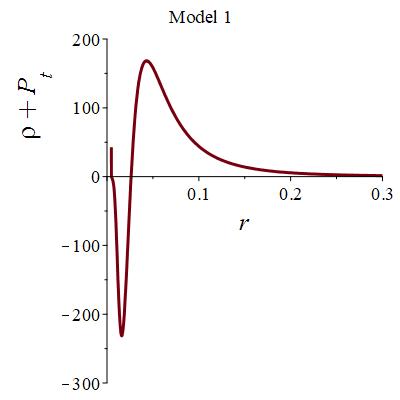}
          \includegraphics[scale=.42]{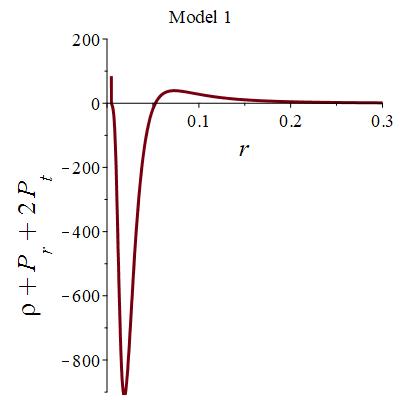}
             \includegraphics[scale=.42]{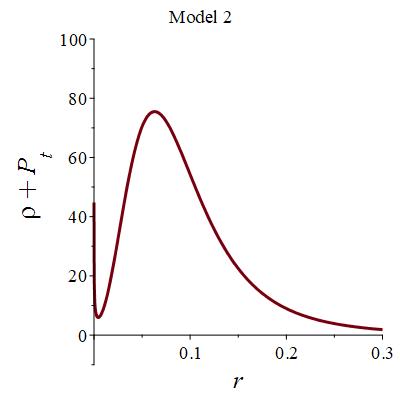}
          \includegraphics[scale=.42]{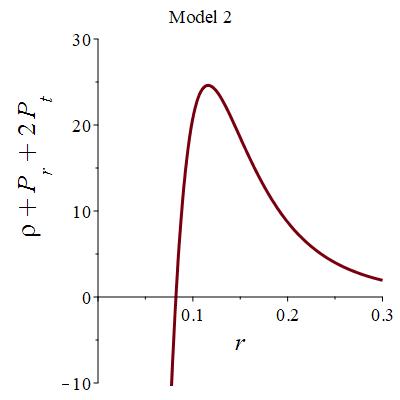}
          \includegraphics[scale=.42]{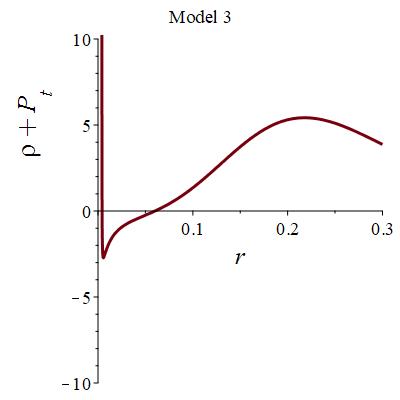}
           \includegraphics[scale=.42]{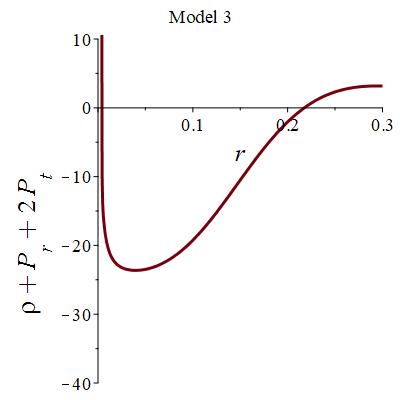}
              \includegraphics[scale=.42]{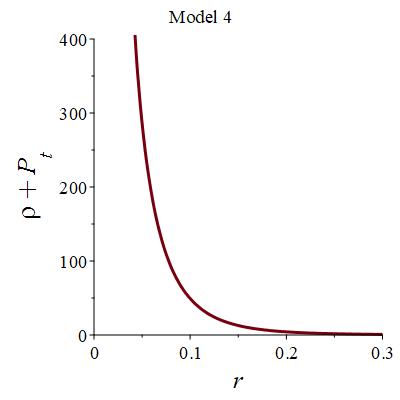}
           \includegraphics[scale=.42]{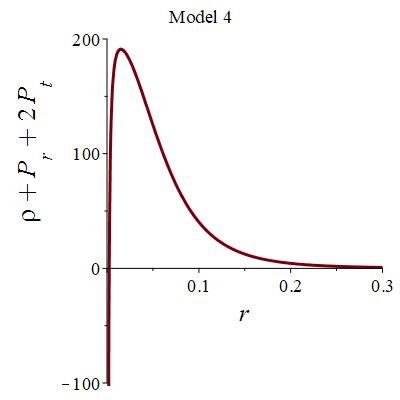}
       \caption{The plot of $\rho(r)+P_t(r)$ and $\rho(r)+P_r(r)+2P_t(r)$ for all models. We have set $l_0=0.1, M=1, a_M=0.01,$ and $Q=0.1$. }\label{f2}
\end{figure*}

\subsection{Black holes in de Sitter space and vanishing charge in VEG}
It is quite easy to specialize these solutions for vanishing charge, i.e., $Q=0$ and non-zero cosmological constant. In that case we get the Minkowski core solution, Bardeen/Hayward/Simpson-Visser solutions:
\begin{equation}
    f(r)=1-\frac{2 M}{r} e^{-\frac{l_0}{r^n}}-\xi e^{-\frac{l_0}{2r^n}}\sqrt{1+\frac{ l_0\, n}{r^{n}}}-\frac{\Lambda r^2}{3},
\end{equation}
\begin{equation}
    f(r)=1-\frac{2 M\,r^2}{(r^2+l_0^2)^{3/2}}-\xi \sqrt{\frac{r^3\, (4 l_0^2+r^2)}{(r^2+l_0^2)^{5/2}}}-\frac{\Lambda r^2}{3},
\end{equation}
\begin{equation}
    f(r)=1-\frac{2 M\,r^2}{r^3+l_0^3}-\xi \frac{\sqrt{r^3\,(4l_0^3+r^3)}}{r^3+l_0^3}-\frac{\Lambda r^2}{3},
\end{equation}
\begin{equation}
f(r)=1-\frac{2M}{\sqrt{r^2+l_0^2}}-\xi \sqrt{\frac{r (r^2+2l_0^2) }{(r^2+l_0^2)^{3/2}}}-\frac{\Lambda r^2}{3},
\end{equation}
respectively. Setting $l_0 \to 0$, we get the point like mass solution in a  de-Sitter space 
\begin{eqnarray}
   f(r)=1-\frac{2M}{r}-\xi-\frac{\Lambda r^2}{3}.
\end{eqnarray}

\section{Energy conditions}
In this section, we turn our attention to study  in more details the energy conditions. These energy conditions are  sets of inequalities depending on energy momentum tensor. In our setup, we shall neglect the energy density due to the cosmological constant, hence for the total energy density we have will take $\rho=\rho_B+\rho_D$. From the Einstein field equations it follows that
	\begin{equation}
  \rho=-\frac{1}{8\pi}\left(\frac{r f'(r)+f(r)-1}{r^2} \right),
\end{equation}
	\begin{equation}
  P_r=\frac{1}{8\pi}\left(\frac{r f'(r)+f(r)-1}{r^2} \right),
\end{equation}
and 
	\begin{equation}
  P_t=\frac{1}{8\pi}\left(\frac{r f''(r)+2f'(r)}{2r} \right).
\end{equation}

For the weak energy condition (WEC), i.e. $T_{\mu\nu} U^{\mu}U^{\nu}$, where $U^{\mu}$ is
a timelike vector we have
\begin{eqnarray}\notag \label{EC1}
\rho (r)&\geq& 0,\\
\rho (r)+P_{i}(r)&\geq& 0.
\end{eqnarray}
For the null energy condition (NEC) using $T_{\mu\nu} k^{\mu} k^{\nu}$, where $k^{\mu}$ is null vector we have
\begin{equation}\label{EC2}
\rho (r)+P_{i}(r)\geq 0, ~~i= 1,2,3
\end{equation}

Finally, for the strong energy condition (SEC) we have
\begin{eqnarray}\label{EC3}\notag
\rho (r)+ \sum P_{i}(r) &\geq 0&,\\
\rho (r)+P_{i}(r)&\geq 0&.
\end{eqnarray}
From these conditions, it follows that 
\begin{equation}
\rho (r)+P_t(r)\geq 0,
\end{equation}
and 
\begin{equation}
\rho (r)+P_{r}(r)+2P_t(r)\geq 0.
\end{equation}

In Fig. (\ref{f2}) we present our results for the energy conditions for all models for some specific values of parameters. We see from these plots that the energy conditions in general are not satisfied in the deep core of the black holes. In particular, for given parameter values, we observe that for Model 2 and Model 4, the WEC energy condition is satisfied but not for Model 1 and Model 3. On the other hand, the SEC is violated for all models.  Perhaps, the most interesting case is the Model 4, where we can see that this energy condition is violated only at $r\to 0$.

\section{Generating rotating black hole solutions in VEG }
In this section, we are interested to obtain rotating black hole geometries and, to simplify the work, we shall also neglect the term proportional to the cosmological constant ($\Lambda r^2/3$) since $\Lambda$ is very small. This term is important however only in large scale structures, while for a finite distance contribution, i.e.,  when the observer and the black hole are not very far away, this term can be neglected in the metric.  Let us start from the static and spherically symmetric charged metric given by
\begin{equation}
    ds^2=-f(r) dt^2+\frac{ dr^2}{g(r)}+ h(r)(d\theta^2+\sin^2\theta d\phi^2)
\end{equation}
where $f(r)=g(r)$, and $h(r)=r^2$, for Model 1, Model 2, and Model 3, respectively. In addition, for the Model 4 we have $h(r)=r^2+l_0^2$. Now we can use the modified NJ algorithm introduced by Azreg-A\"{i}nou's and based on the non-complexification procedure (see for details \citep{Azreg-Ainou:2014pra}). First, we have to  transform the metric from Boyer-Lindqiust coordinates $(t, r, \theta, \phi)$ to Eddington-Finkelstein coordinates $(u, r, \theta, \phi)$, then by applying the steps presented in \citep{Azreg-Ainou:2014pra} we can obtain the effective rotating BH metric in Kerr-like coordinates:
	\begin{eqnarray}\label{eq:rotmet}\notag
	ds^2 &=& \frac{H}{\Sigma} \Big[-\frac{\Delta}{\Sigma}(dt-a\sin^2\theta d\phi)^2+\frac{\Sigma}{\Delta}dr^2+\Sigma d\theta^2 \\
	&+&\frac{\sin^2\theta}{\Sigma}(a dt-(k+a^2)d\phi)^2 \Big],  
	\end{eqnarray}
where $a$ is the specific angular momentum and the following definitions have been used:
	\begin{eqnarray}
	\Delta(r)&=&g(r)h(r)+a^2,\\
	k(r)&=&\sqrt{g(r)/f(r)}\,h(r)\\
		\Sigma(r)&=&k+a^2\cos^2\theta,
	\end{eqnarray}
	
	It is also important to say that the function $H(r, \theta, a)$ is still arbitrary at this point we can use the condition from the cross-term from the Einstein tensor, i.e., $G_{r \theta}=0$, to obtain physically acceptable rotating solutions. This constraint yields the following differential equation \citep{Azreg-Ainou:2014pra}
\begin{eqnarray}
(k(r)+a^2 y^2)^2(3 H_{,r}H_{,y^2}-2H H_{,r y^2})=3 a^2 k_{,r}H^2,
\end{eqnarray}
with $y\equiv\cos \theta$. The above equation has the following solution
\begin{eqnarray}\label{eq:hfunc}
H\equiv \sqrt{g(r)/f(r)} \,k+a^2\cos^2\theta.
\end{eqnarray}
Therefore, by using the above relations, we can obtain the following corresponding solutions in VEG:
	\begin{itemize}
	    \item Model 1
	\end{itemize}
	\begin{eqnarray}\notag
    \Delta(r)&=&r^2-\left(2Mr-Q^2\right)e^{-\frac{l_0}{r^n}}-2r\sqrt{a_M} e^{-\frac{l_0}{2r^n}} \mathcal{A}(r)+a^2,\\
  k(r)&=&r^2,\,\,\,\text{and}\,\,\,\,\Sigma(r)=H(r)= r^2+a^2\cos^2\theta,
\end{eqnarray}
 where $\mathcal{A}(r)$ is given by Eq. (11). When the charge  vanishes, i.e., $Q=0$, we get
 \begin{equation}
     \Delta(r)=r^2-2Mre^{-\frac{l_0}{r^n}}-\xi r^2 e^{-\frac{l_0}{2r^n}}\sqrt{1+\frac{ l_0\, n}{r^{n}}}+a^2.
 \end{equation}
	\begin{itemize}
	    \item Model 2
	\end{itemize}
\begin{eqnarray}\notag
     \Delta(r)&=&r^2-\frac{2M r^4}{\left(r^2+l_0^2\right)^{3/2}}+\frac{Q^2 r^4\mathcal{H}}{(r^2+l_0^2)^2}-2r^2\sqrt{a_M\mathcal{G}}+a^2,\\
      k(r)&=&r^2,\,\,\,\text{and}\,\,\,\,\Sigma(r)=H(r)= r^2+a^2\cos^2\theta,
\end{eqnarray} 
 where $\mathcal{H}(r)$  and $\mathcal{G}(r)$ are given by Eqs. (19)-(20). When $Q=0$, we get
 \begin{equation}
     \Delta(r)=r^2-\frac{2M r^4}{\left(r^2+l_0^2\right)^{3/2}}-\xi r^2 \sqrt{\frac{r^3\, (4 l_0^2+r^2)}{(r^2+l_0^2)^{5/2}}}+a^2.
 \end{equation}

	\begin{itemize}
	    \item Model 3
	\end{itemize}
\begin{eqnarray}\notag
     \Delta(r)&=&r^2-\frac{(2Mr-Q^2)r^4}{r^4+(2Mr+Q^2)l_0^2}- \frac{2r^2\sqrt{a_M \mathcal{K}}}{r^4+(2Mr+Q^2)l_0^2}+a^2,
     \end{eqnarray}
     \begin{eqnarray}\notag
      k(r)&=&r^2,\,\,\,\text{and}\,\,\,\,\Sigma(r)=H(r)= r^2+a^2\cos^2\theta,
\end{eqnarray}
 where $\mathcal{K}(r)$ is given by Eq. (28). By setting $Q=0$, and rescaling the parameters $2M l_0^2 \to l_0^3$, we get
 \begin{equation}
     \Delta(r)=r^2-\frac{2 M\,r^4}{r^3+l_0^3}-\xi r^2 \frac{\sqrt{r^3\,(4l_0^3+r^3)}}{r^3+l_0^3}+a^2
 \end{equation}
\begin{itemize}
	    \item Model 4
	\end{itemize}	
	\begin{eqnarray}\notag
	    \Delta(r)&=&r^2+l_0^2-2M \sqrt{r^2+l_0^2}+Q^2-2 \sqrt{ \frac{a_M r \mathcal{R}}{(r^2+l_0^2)^{1/2}}}+a^2,
	    \end{eqnarray}
	    \begin{equation}
	    k(r)=r^2+l_0^2,\,\,\,\text{and}\,\,\,\,\Sigma(r)=H(r)= r^2+l_0^2+a^2\cos^2\theta,
	\end{equation}
    where $\mathcal{R}(r)$ is given by Eq. (35). Finally for $Q=0$, we can obtain 
    \begin{equation}
        \Delta(r)=r^2+l_0^2-2M \sqrt{r^2+l_0^2}-\xi \sqrt{r(r^2+l_0^2)^{1/2}(r^2+2l_0^2)}+a^2,
    \end{equation}
    	and we recall that $\xi=2 \sqrt{a_M M}$. In the special limit, we obtain the point like mass profile when $Q\to 0$, and $l_0 \to 0$. In fact, all these solutions again coincide
 \begin{eqnarray}
     \Delta(r)=r^2-2Mr-\xi r^2+a^2,
 \end{eqnarray}
and, again, we get the correspondence with the rotating Kerr black hole with a global monopole or cloud of string [we can identify $\xi=8\pi \eta^2$, see for more details \cite{Haroon:2019new}].  We would like to comment that the above solutions are in general non-asymptotically flat spacetimes, therefore, there is no guarantee that the NJAA will lead to exact solutions. It has been shown that for a spherical symmetric solution with the modified Newman-Janis algorithm \cite{Azreg-Ainou:2014pra} one can solve  the Einstein field equations with the energy-momentum tensor represented by a properly chosen tetrad of the vector given by $T^{\mu \nu}=e^{\mu}_{a}e^{\nu}_{b}T^{ab}$, where $T^{ab}=(\rho,P_r,P_{\theta},P_{\phi})$, with $\rho=\rho_B+\rho_D$ in our case. One has to find an orthogonal basis with the energy momentum tensor components are given as
\begin{eqnarray}\nonumber
\rho &=&\frac{1}{8\pi}\emph{e}^\mu_t\,\emph{e}^\nu_t \,\emph{G}_{\mu\nu},\quad
P_r =\frac{1}{8\pi}\emph{e}^\mu_r\,\emph{e}^\nu_r \,\emph{G}_{\mu\nu},\\\label{m1}
P_\theta &=&\frac{1}{8\pi}\emph{e}^\mu_\theta\,\emph{e}^\nu_\theta \,\emph{G}_{\mu\nu},\quad
P_\varphi =\frac{1}{8\pi}\emph{e}^\mu_\phi\, \emph{e}^\nu_\phi\, \emph{G}_{\mu\nu}.
\end{eqnarray}

In other words, we need an orthogonal bases such that the Einstein field equations are satisfied $G_{\mu \nu}= 8\pi T_{\mu \nu}$. It has been shown that indeed such a bases exists and one such orthogonal bases is the following choice \cite{Azreg-Ainou:2014pra}
\begin{eqnarray}\label{basis}
{\emph{e}}^\mu_t&=&\frac{1}{\sqrt{\Sigma \Delta}}\left(r^2+a^2,0,0,a\right),\quad
\emph{e}^\mu_r=\frac{\sqrt{\Delta}}{\sqrt{\Sigma}}\left(0,1,0,0\right),\\\nonumber
\emph{e}^\mu_\theta&=&\frac{1}{\sqrt{\Sigma}}\left(0,0,1,0\right),\quad
\emph{e}^\mu_\phi=\frac{1}{\sqrt{\Sigma} \sin\theta}\left(a \sin^2\theta,0,0,1\right).
\end{eqnarray}

Due to the economy of space, we done give here the full expressions for the Einstein tensor components, but the interested reader can see \cite{Azreg-Ainou:2014pra} and \cite{Jusufi:2019nrn} for details. For the momentum tensor components we find
\begin{eqnarray}
\rho &=&\frac{2\Upsilon'(r)r^2}{8 \pi \Sigma^2}=-P_r,\\
P_{\theta}&=&P_{\phi}=P_r-\frac{\Upsilon''(r)r+2\Upsilon'(r)}{8 \pi \Sigma}.
\end{eqnarray}
where $\Upsilon(r)=r(1-f(r))/2$. Thus, we can say that, the rotating solutions presented in this work solve the Einstein field equations provided the energy momentum tensor components are given by Eqs. (74)-(75). We interpret the black hole solutions as effective rotating geometries which can be used by assuming that the black hole and the observer are located at some finite distance between them. On the other hand, for specific parameter values, the spacetime may have different pathologies. Some of these conditions are: the metric determinant must always be negative, $g_{\phi\phi}$ is always greater then $0$, and $g_{rr}$ remains finite. Imposing these conditions on the metric we get the constraint $ \Delta(r)/H(r) > 0.$ Finally, we are not going to study the phenomenological aspect of these solutions in the present draft, these solutions however, can open new window to test VEG in the strong gravity regime. 

\section{Conclusions}
In this paper we used the relation between the apparent dark matter and the baryonic matter and we find charged and regular black holes in VEG. We considered four different models for the mass profile, such as black holes with asymptotically Minkowski core, T-duality, Frolov and Simpson-Visser type solutions, as well as the Bardeen and Hayward type solutions in VEG. Although, the central singularity is not removed for the Bardeen and Hayward solution in VEG, we argued that from a point of view of a free falling observer, the light/particles never reach the central singularity. Instead, the plot for the velocity for light/particles can reach a minimal value and then rapidly increases for small value of $r$. Such a singularity is known as a timelike singularity. We found that apparent dark matter leads to non-asymptotically flat spacetime geometry and, in the special limit of a point like mass distribution, the global monopole-like solution with a deficit angle is obtained. From the analyses of the energy conditions we found that in general the SEC is violated at the deep core of the black hole. In the final part we used the modified Newman--Janis-Azreg-A\"inou algorithm to obtain the corresponding effective rotating metrics. In the near future, we are planing to study the phenomenological aspect of black holes in VEG, including light deflection in the strong and weak gravity regime, the stability and quasinormal modes, thermodynamics, shadow images and particle dynamics. 

\end{document}